\documentclass[a4paper]{jpconf}

\usepackage{graphicx}
\usepackage{epstopdf}

\usepackage{natbib} 
\bibpunct{(}{)}{;}{a}{}{,} 
\usepackage{iopams}

\def\apj{ApJ}%
\def\apjl{ApJ}%
\def\aap{A\&A}%
\def\mnras{MNRAS}%
\def\solphys{Sol.~Phys.}%
\def\nat{Nature}%
\begin{document}

\title{Magnetic tornadoes and chromospheric swirls -- Definition and classification.}
\author{Sven Wedemeyer$^{1}$, 
Eamon Scullion$^{1}$, 
Oskar Steiner$^{2}$,  
Jaime de la Cruz Rodriguez$^{3}$
and 
Luc Rouppe van der Voort$^{1}$} 
  
\address{$^1$ Institute of Theoretical Astrophysics, University of Oslo, 
P.O. Box 1029 Blindern, N-0315~Oslo, Norway}
\address{$^2$ Kiepenheuer Institute for Solar Physics, Sch{\"o}neckstr. 6-7 D-79104 Freiburg, Germany}
\address{$^3$ Department of Physics and Astronomy, Uppsala University, Box 516, SE-75120 Uppsala, Sweden}

\ead{sven.wedemeyer@astro.uio.no}

%%%%%%%%%%%%%%%%%%%%%%%%%%%%%%%%%%%%%%%%%%%%%%%%%%%%%%%%%%%%%%%%%%%%%%%%%%%%%%%%%%%%%%
\begin{abstract}
Chromospheric swirls are the observational signatures of 
rotating magnetic field structures in the solar atmosphere, also known 
as magnetic tornadoes.  
Swirls appear as dark rotating features in the core of the spectral line 
of singly ionized calcium at a wavelength of 854.2~nm.
This signature can be very subtle and difficult to detect given the 
dynamic changes in the solar chromosphere. 
Important steps towards a systematic and objective detection method are the 
compilation  and characterization of a statistically significant sample of 
observed and simulated chromospheric swirls.  
Here, we provide a more exact definition of the chromospheric swirl
phenomenon and also present a first morphological classification of swirls 
with three types: (I)~Ring, (II)~Split, (III)~Spiral.  
We also discuss the nature of the magnetic field structures connected to 
tornadoes and the influence of limited spatial resolution on the appearance of 
their photospheric footpoints.
\end{abstract}

%%%%%%%%%%%%%%%%%%%%%%%%%%%%%%%%%%%%%%%%%%%%%%%%%%%%%%%%%%%%%%%%%%%%%%%%%%%%%%%%%%%%%%
%%%%%%%%%%%%%%%%%%%%%%%%%%%%%%%%%%%%%%%%%%%%%%%%%%%%%%%%%%%%%%%%%%%%%%%%%%%%%%%%%%%%%%
\section{Introduction} 

\begin{figure}[t!]
\begin{center}
\includegraphics[width=9.4cm]{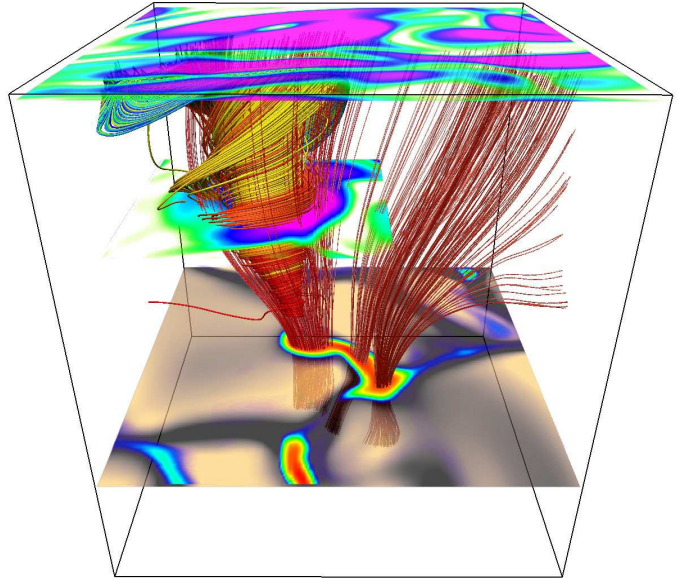}
\vspace*{-3mm}
\caption{Visualization of a close-up region from the numerical simulation with  
a magnetic tornado.  
The red (mostly vertical) lines represent magnetic field lines, whereas the 
red/orange/yellow wound lines trace the velocity field in the tornado in this snapshot. 
The lower surface shows the granulation superposed with the absolute magnetic field 
strength at the bottom of the photosphere. 
The other layers show the horizontal velocity at heights of 1000\,km and 2000\,km 
with pink marking the highest speeds. 
Created with VAPOR \citep{vapor_clyne2007}. 
\label{fig:vapor}
} 
\end{center}
\end{figure}

The combination of photospheric vortex flows and magnetic fields produces 
rotating magnetic field structures in the solar atmosphere, which are both 
seen in high-resolution observations and numerical simulations  
\citep[][ hereafter Paper~I]{2012Natur.486..505W}. 
The streamlines, which trace the simulated velocity field of the plasma in the 
rotating field structures, have a narrow footpoint in the photosphere, which 
broadens into a wide funnel in the chromosphere above (see Fig.~\ref{fig:vapor}). 
This appearance is reminiscent of tornadoes on Earth, which led to the name 
`magnetic tornadoes' for this solar phenomenon. 
The term `tornado' has already been used before for vertically aligned spiral 
structures in connection with prominences 
\citep[see Chap.~10 by Tandberg-Hansen in][]{1977ASSL...69.....B}. 
Such `solar tornadoes'\footnote{See the ESA press release at http://www.esa.int/esaCP/Pr\_15\_1998\_i\_EN.html}  were  observed  
with the Coronal Diagnostic Spectrometer \citep[CDS,][]{1995SoPh..162..233H}
onboard the Solar and Heliospheric Observatory \citep[SOHO,][]{1995SoPh..162....1D}, 
implying that rotation may play an important role for the dynamics of the solar transition region   
\citep{1997SoPh..175..457P,1998SoPh..182..333P,2000A&A...355.1152B}.  
More recent observations with the Solar Dynamics Observatory \citep[SDO,][]{2012SoPh..275...17L} 
revealed more details of this  phenomenon
\citep{2012ApJ...752L..22L}.
There is some indication that these giant tornadoes can be explained 
as rotating magnetic structures driven by underlying photospheric vortex flows just 
like the magnetic tornadoes described in Paper~I   
\citep{2012ApJ...756L..41S, 2012ApJ...761L..25O}. 
The connection between the prominence-related giant tornadoes and the smaller 
magnetic tornadoes is not clear yet. 
However, it is established that photospheric vortex flows, which are an essential 
ingredient of tornadoes, are a common phenomenon on the Sun. 
Vortex flows are seen both in 
observations 
\citep[e.g.][]{1988Natur.335..238B, 2008ApJ...687L.131B, 2010ApJ...723L.139B, 2010ApJ...723L.180S, 2011MNRAS.416..148V}
and in numerical simulations 
\citep[e.g.][ Paper~I]{1985SoPh..100..209N, 1998ApJ...499..914S,2005A&A...429..335V, 2010ApJ...723L.180S,  2011ApJ...727L..50K,2011A&A...526A...5S,2012A&A...541A..68M}.

Here, we present a first definition and classification of chromospheric 
swirls \citep[][ hereafter Paper~II]{2009A&A...507L...9W}, 
which is the essential observational signature of magnetic tornadoes, 
and address the appearance of the magnetic footpoints, which are another essential 
indicator of a magnetic tornado.

% --------------------------------------------------------------------------------
% --------------------------------------------------------------------------------

% ================================================================================
\section{Numerical simulations}
\label{sec:sim}

% --------------------------------------------------------------------------------
\begin{figure}[t!]
\begin{center}
\vspace*{-2mm}
\includegraphics[width=10.9cm]{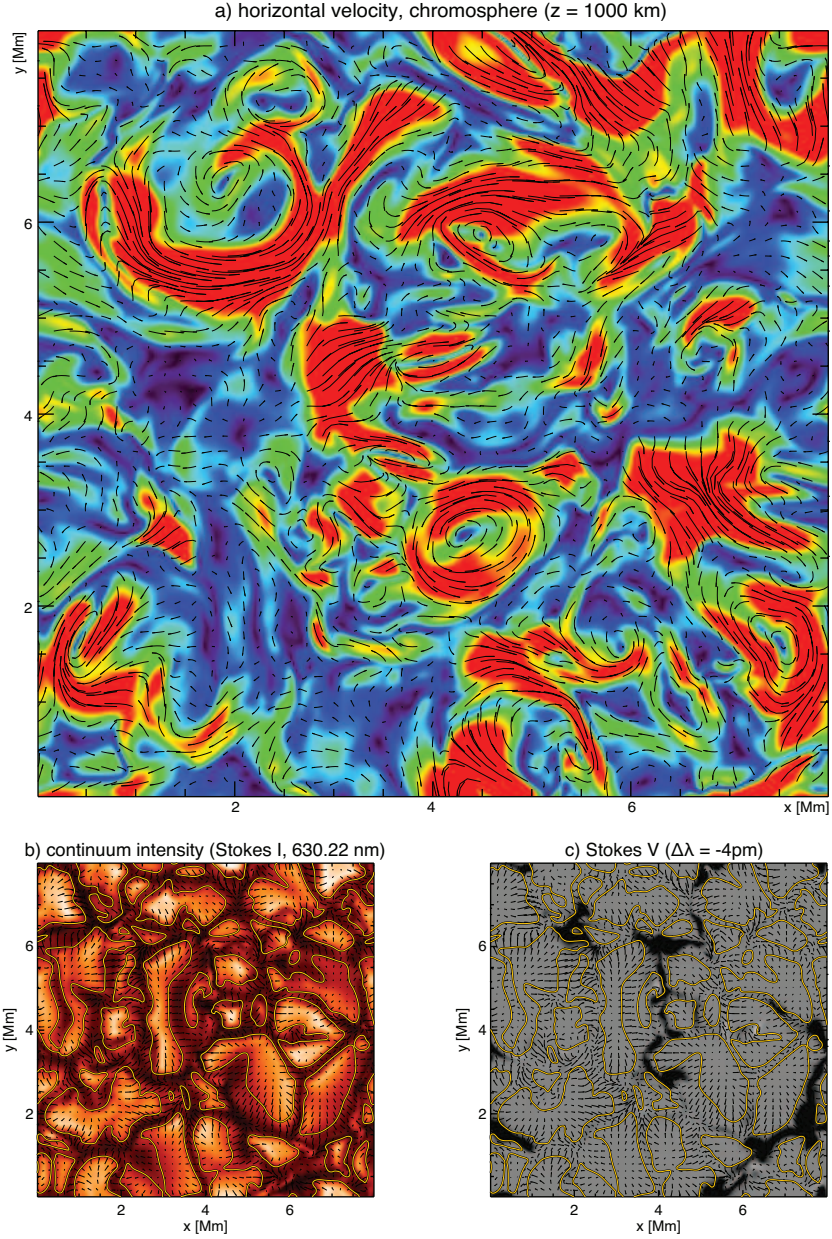}
\vspace*{-2mm}
\caption{\textbf{(a)}~Horizontal cross-section through a simulation snapshot at 
$z = 1000$\,km (chromosphere) displaying in colors the absolute value of the 
horizontal velocity and the projected streamlines. 
\textbf{(b)}~The corresponding continuum intensity (Stokes~I at $\lambda = 630.22$\,nm) 
and 
\textbf{(c)}~Stokes~V at $\Delta \lambda = -4$\,pm from line center.
The contours for $v_z = 0$\,km/s in the lower panels outline the granule boundaries, 
while the streamlines trace the horizontal velocity field. 
\label{fig:vhor}
} 
\end{center}
\end{figure}
% --------------------------------------------------------------------------------

The analysis in Paper~I was primarily based on numerical simulations with the 3-D 
radiation magnetohydrodynamics code \mbox{CO$^5$BOLD} 
\citep{2012JCoPh.231..919F,2004A&A...414.1121W}. 
The computational box of these models has a horizontal size of 8\,Mm\,$\times$\,8\,Mm 
and extends vertically from 2.4\,Mm below the optical depth level 
$\tau_c = 1$ to 2.0\,Mm above it, i.e., to the top of the chromosphere.
The initial model was derived from a non-magnetic simulation, which was supplemented 
with an initially vertical, homogeneous magnetic field with a field strength of 
$B_0 = 50$\,G.
Periodic lateral boundaries and an open lower boundary were used, whereas the 
top boundary is transmitting for hydrodynamics and outward radiation. 
The tangential magnetic field component vanishes at the top boundary, i.e., there,  
the magnetic field is vertical.\linebreak  
No artificial limiting of the Alfv{\'e}n speed has been imposed.
The analysis of the magnetic tornadoes in Paper~I was done on a sequence 
with 1\,s cadence. 
In Fig.~\ref{fig:vhor}, cross-sections through one of these snapshots are shown. 
The upper panel displays the (horizontal) flow pattern in the chromosphere. 
There are about a dozen chromospheric swirls of different types and sizes 
(some cases more obvious than others), indicating that  
swirl events are very common in the simulations. 
Typically, every second magnetic footpoint in the photosphere is connected 
to a swirl in the chromosphere above.

% ================================================================================
\section{Magnetic tornadoes and chromospheric swirls -- Definition and classification} 

% ================================================================================
\subsection{Conditions for the formation of tornadoes}

In the photosphere, the plasma flows away from the granule interiors and 
towards the intergranular lanes, where the cooled plasma shoots back down into the 
convection zone (see Fig.~\ref{fig:vhor}b). 
The flow carries a net angular momentum. 
Due to the conservation of angular momentum, the downflows in the lanes 
can create vortex flows 
\citep[`bathtub effect', see, e.g.,][]{1985SoPh..100..209N}. 
Vortex flows are formed most easily at the vertices of lanes, where plasma 
meets from the neighbouring granules. 
The flows also advect magnetic field into the intergranular lanes, where the 
field is concentrated and thus amplified. 
Consequently, vortex flows and magnetic field concentrations often exist 
at the same locations. 
A stationary magnetic tornado is generated when the magnetic field 
concentration is perfectly co-located with the vortex flow, resulting in the rotation 
of the entire magnetic structure. 
However, exact co-location is not always perfectly fulfilled. 
A magnetic field structure can be partially pushed into a vortex and out again, 
which results in a partial rotation of the magnetic field structure but not in a 
stationary tornado. 
We therefore expect tornadoes to exhibit a spectrum of rotational behavior, 
from partial to stationary rotation.

% ================================================================================
\subsection{Chromospheric swirls as observable signature of tornadoes} 

Swirls have been discovered as dark rings in the line core of 
the calcium infrared triplet line at a wavelength of 854.2\,nm
during an observation campaign at the Swedish 1-m Solar Telescope (SST) \citep{2003SPIE.4853..341S}
in 2008 (Paper~II). 
We refer to this observational imprint as the 
Ca~line core signature (CLCS) or Ca~854~swirl hereafter.
It is the only direct indicator of chromospheric swirls and magnetic tornadoes 
known so far.
This is due to the fact that 
detecting swirls is a challenging task, which requires stable, high-quality 
image series with high spatial and temporal resolution for a purely chromospheric 
diagnostic. 
In this respect, a narrow wavelength transmission range is crucial 
because the subtle swirl signature is only visible if the
intensity signal is not contaminated by large contributions from the line wings, 
which are formed lower in the atmosphere. 
The CRISP instrument \citep{2008ApJ...689L..69S}  at the SST offers 
this possibility, but even with this instrument, the observation of 
chromospheric swirls is still challenging. 

% ================================================================================
\subsection{Definition -- What is a chromospheric swirl?}

\begin{figure}[t]
\begin{center}
\includegraphics[width=10.2cm]{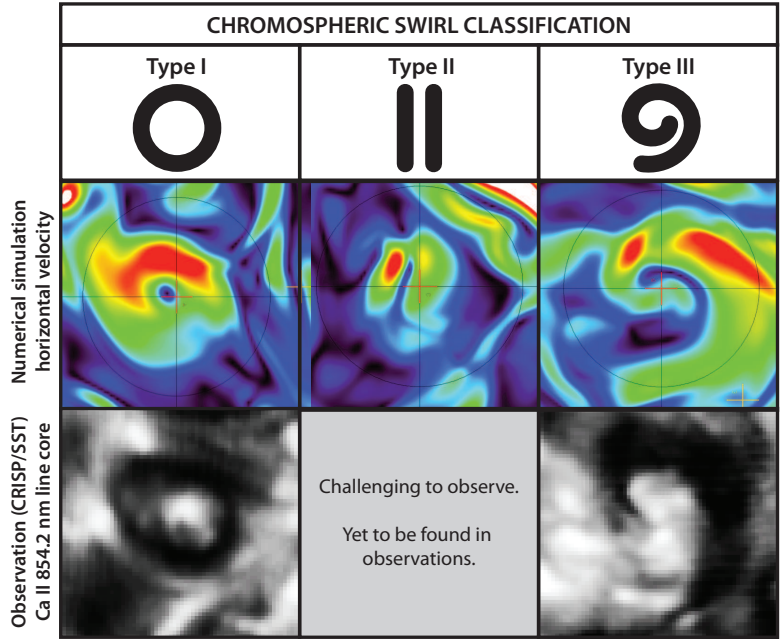}
\caption{The three major types of chromospheric swirls: 
Type~I (Ring), type~II (Split), and type~III~(Spiral).  
The middle row shows the color-coded horizontal velocity in horizontal 
cross-sections at $z = 1000$\,km for swirls in the numerical simulation (Paper~I). 
Observed examples of type~I and III swirls are presented in the bottom row. 
The images are taken in the line core of the Ca\,II infrared triplet line at 
854.2\,nm (Paper~II). 
\label{fig:types}
}
\end{center}
\end{figure}

In accordance with Paper~II, we note that the swirl pattern can be seen as darkening in the  
Ca\,II\,854.2\,nm line core at wavelengths down to $\Delta \lambda \approx 70$\,pm. 
The intensity at wavelengths further away from the line core originate from deeper down 
in the atmosphere, where the rotating magnetic structure has a smaller extent and is 
thus less clearly visible. 
There is a corresponding brightening with mostly the same pattern in the red line 
wing close to the line core, produced by the Doppler shift within the swirl. 
Based on these observations and supporting numerical simulations, 
we formulate the following conditions that need to be fulfilled for a chromospheric swirl: 
\begin{enumerate} 
\item Ca\,II\,854\,nm line core images exhibit a dark ring, 
a ring fragment or a spiral (see Fig.~\ref{fig:types}). 
\item The Ca~854~swirl is rotating.
\item The Ca~854~swirl is visible for several minutes.
\item Vertical velocities (Doppler-shifts at disk centre) on the order of 
at least 2\,km/s and more can be measured at locations coinciding with the Ca~854~swirl.
\item A photospheric magnetic field concentration is visible at the same location, e.g., in 
the form of a magnetic bright point and/or in the magnetogram. 
\end{enumerate}

% --------------------------------------------------------------------------------
% --------------------------------------------------------------------------------

% ================================================================================
\subsection{Classification} 

Swirls can have different shapes in the Ca line core images as already stated in 
Paper~II. 
The shapes are also seen in horizontal cross-sections of the horizontal 
velocity at chromospheric heights in the numerical models.
The three most prominent types are shown in Fig.~\ref{fig:types}: 

\begin{center}
\begin{tabular}{lll}
Type~I:  &Ring.         &Example in Fig.~\ref{fig:vhor}a: $[x = 4.3$\,Mm, $y = 2.7$\,Mm$]$  \\
Type~II: &Split.        &Example in Fig.~\ref{fig:vhor}a: $[x = 4.0$\,Mm, $y = 4.6$\,Mm$]$    \\ 
Type~III:&Spiral.\qquad &Example in Fig.~\ref{fig:vhor}a: $[x = 1.9$\,Mm, $y = 6.4$\,Mm$]$    \\
\end{tabular}
\end{center}

\noindent All three types are seen in numerical simulations and types~I and III are observed. 
A type~I swirl is not necessarily a closed ring but can consist of ring fragments, too. 
Some simulation examples suggest that type~I swirls can evolve into type~II as 
the circular flow pattern becomes increasingly elongated until the swirl signature 
fades away. 
A type~III swirl can also have multiple spiral arms. 
Based on a preliminary analysis of the simulations, half of the cases 
are classified as type~I, and one fourth as type~II 
and III, respectively. 
Type~I swirls are possibly connected to open field structures, which can rotate more 
freely than the legs of closed loops. 
The latter build up more twist, which might result in spiral-like  
type~III swirls.

%%%%%%%%%%%%%%%%%%%%%%%%%%%%%%%%%%%%%%%%%%%%%%%%%%%%%%%%%%%%%%%%%%%%%%%%%%%%%%%%%%%%%%
\section{Photospheric footpoints of magnetic tornadoes}
%%%%%%%%%%%%%%%%%%%%%%%%%%%%%%%%%%%%%%%%%%%%%%%%%%%%%%%%%%%%%%%%%%%%%%%%%%%%%%%%%%%%%%
% --------------------------------------------------------------------------------
\begin{figure}[t!]
\begin{center}
\includegraphics[width=15cm]{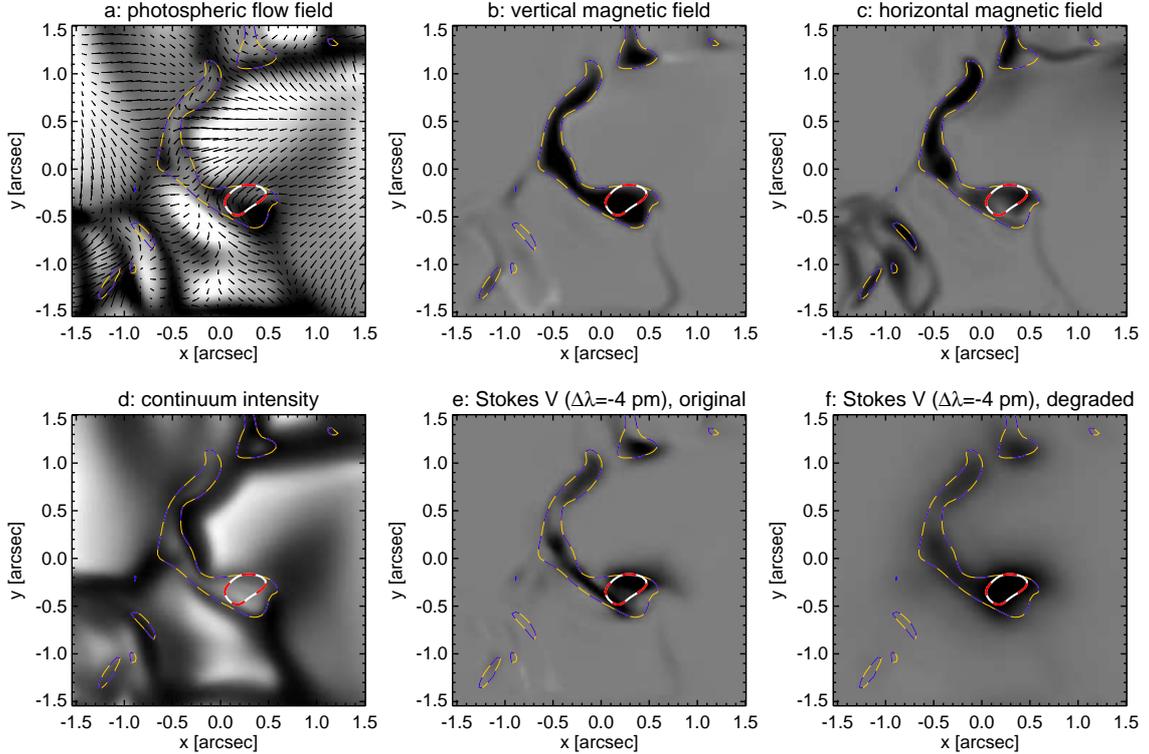}
\caption{Effect of limited spatial resolution on the appearance of the magnetic 
flux topology shown for a small close-up region from a magnetohydrodynamical 
simulation (Fig.~\ref{fig:vhor}b-c, Paper~I). 
\textbf{a)}~Flow field at the bottom of the photosphere. 
The vertical velocity component is plotted in grey-scale, whereas the horizontal 
components in the plane are represented by arrows.  
\textbf{b)}~The vertical magnetic field component and 
\textbf{c)}~the horizontal magnetic field ($({B_x^2+B_y^2})^{1/2}$) in the same plane. 
\textbf{d)}~The continuum intensity (Stokes~I at a wavelength of $\lambda = 630.22$\,nm).
\textbf{e)}~Original Stokes~V map at $\Delta \lambda = -4$\,pm from line center and 
\textbf{f)}~the Stokes~V map after a PSF has been applied.
The blue-yellow dashed lines are contours of the absolute magnetic field strength at 
$|B| = 400$\,G, whereas the red-white dashed contour outlines the magnetic 
concentration that is visible in the degraded image~(f).
\label{fig:magnetogram}
} 
\end{center}
\end{figure}
% --------------------------------------------------------------------------------

The existence of a photospheric bright point as magnetic field indicator 
below a chromospheric swirl is a 
necessary condition for the detection of a magnetic tornado. 
However, the sizes of magnetic bright points (MBP) can be very small so that 
their identification and tracking over time is affected by the spatial 
resolution of the employed instrument. 
In the following, we discuss implications for the detection of tornado-related 
MBPs.

\paragraph{The magnetized plasma of the solar atmosphere -- A continuous medium.}
Starting from the initial model, the magnetic field quickly concentrates
in the vertices of intergranular lanes and a complex and 
entangled magnetic field structure evolves. 
It is possible to identify groups of magnetic field lines that rotate and/or sway 
together. 
These groups usually do not persist for long because the magnetic field is continuously 
rearranged on short time scales connected to the lifetime of granules and the 
resulting changes in the photospheric flow field. 
Like other authors before \citep[e.g.,][]{2006ApJ...642.1246S}, 
we argue that the appearance of flux tubes is (at least partially) the product of the 
limited spatial resolution of the observations, resulting in  
apparently separated MBPs. 
It would be more appropriate to talk of a continuous magneto-fluid in which 
the field lines at neighbouring locations exhibit similar dynamics, rather 
than to describe the magnetic field as individual, isolated flux tubes.

The effect of limited spatial resolution is illustrated in Fig.~\ref{fig:magnetogram}, 
which displays a close-up region in a numerical simulation snapshot 
(see Sect.~\ref{sec:sim}). 
As result of the convective flows (Fig.~\ref{fig:magnetogram}a), the magnetic field 
is arranged in the form of continuous sheets and knots 
(see $B_z$ in panel~b and $B_h = ({B_x^2+B_y^2})^{1/2}$ in panel~c).
The strongest field concentrations appear as bright features in the continuum 
intensity (panel~d), although the correlation between magnetic field strength and 
intensity is not strict. 
The corresponding magnetogram for the Fe\,I line at a wavelength of 630.2\,nm 
is shown in panel~e. 
This magnetogram, i.e., the Stokes~V signal, is calculated with the full-Stokes radiative transfer code NICOLE \citep{2011A&A...529A..37S,2012A&A...543A..34D}
along vertical lines of sight. 
The Stokes signal is formed over a height range in the atmosphere, whereas  
the magnetic field in panels~b and c refers to a single horizontal 
cross-section at the bottom of the photosphere only. 
The resulting magnetogram  appears therefore more intermittent and less continuous than 
the magnetic field in the horizontal plane. 
Next, the magnetogram is convolved with a point spread function
\citep[PSF, see][]{2008A&A...487..399W} 
that simulates the effect of a telescope with an aperture of 50\,cm, which is equivalent to the 
Solar Optical Telescope 
\citep[SOT,][]{2008SoPh..249..167T,2008SoPh..249..197S,2008ASPC..397....5I,2008SoPh..249..221S}
onboard the Hinode spacecraft \citep{2007SoPh..243....3K}. 
The resulting degraded magnetogram is shown in Fig.~\ref{fig:magnetogram}f. 
It should be noted that this degradation is optimistic. 
In practice, the derivation of the Stokes components from measurements with limited 
spatial, temporal, and spectral resolution would result in magnetograms that are 
even more diffuse and less reliable than the synthetic example presented here. 
Nevertheless, the effect of the limited spatial resolution is obvious. 
Only individual blobs of enhanced Stokes~V signal remain instead of the continuous 
magnetic flux sheet  in the original model atmosphere.  
When observed, these blobs would be identified as MBPs. 
The most prominent MBP is outlined with a red-white dashed contour in Fig.~\ref{fig:magnetogram}.  
Most of the remaining magnetic field (blue-yellow contours) would remain undetected. 
In particular, the thin magnetic flux sheets in the intergranular lanes are 
barely visible.  
In view of this effect, the questions arises how uniquely a MBP can be identified. 
During an observation run, a MBP can be tracked from its first appearance until its disappearance 
as long as it remains sufficiently separated from other MBPs. 
In that case a lifetime can be determined.  
However, the numerical simulations suggest that the magnetic field is 
continously reorganised. 
Some of the magnetic field, which would be observed as part of the MBP, may be advected 
into a neighbouring sheet and, in return, additional field may be advected into the MBP. 
These changes would remain undetected, whereas the same MBP appears to persist. 
The lifetimes of observed MBPs are consistent with the timescales on which 
the granulation pattern evolves \citep[e.g.,][ and references therein]{2013A&A...549A.116J} 
but  the magnetic field (incl. individual MBPs) can nevertheless last beyond the lifetime
of the surrounding granules.

% --------------------------------------------------------------------------------
\section{Discussion and Conclusions} 

The primary way to detect a chromospheric swirl and thus a magnetic tornado 
is the Ca\,II\,854.2\,nm line core signature (CLCS) so far, which requires a 
narrow filter transmission range. 
These requirements make it challenging to observe chromospheric swirls 
and, to our knowledge, only CRISP/SST observations have repeatedly produced successful 
swirl detections so far. 
Ultimately, more observations, preferably with a variety of instruments, are 
needed in order to gather a statistically significant sample that allows 
for a detailed characterization of magnetic tornadoes, including their 
occurrence, distribution of sizes and lifetimes  and the related energy transport rates.  
These small-scale events represent a challenging test for the 
next generation of solar telescopes like ATST \citep{2011ASPC..437..319K}
and EST \citep{2010AN....331..615C}. 
High spatial resolution is not only important for the observations but also for the numerical 
simulations. 
The successful modelling of magnetic tornadoes requires the sufficient resolution of 
the narrow intergranular lanes, where photospheric vortex flows and magnetic fields 
coalesce. 

% --------------------------------------------------------------------------------
\ack SW thanks the organizers of the conference 
``Eclipse on the Coral Sea: Cycle 24 Ascending'' 
GONG 2012/LWS/SDO-5/SOHO 27, which has been held in Palm Cove, Australia, in 
 2012. 

% --------------------------------------------------------------------------------
\small
\bibliographystyle{aa} 

%\bibliography{swb,mdwarf}

% --------------------------------------------------------------------------------
\end{document}